\documentclass[a4paper, 10 pt, conference]{ieeeconf} 
\IEEEoverridecommandlockouts
\usepackage[top=54pt, bottom=104pt,left=54pt,right=37pt]{geometry}

\usepackage{amsfonts,amsmath,amssymb} % assumes amsmath package installed

\newcommand{\dif}{\,\mathrm{d}}		  % Define dif
\DeclareMathOperator{\diag}{diag}         % Define diag operator
\usepackage{psfrag,color}
\usepackage{enumerate,cite,latexsym,graphicx}
\newtheorem{theorem}{Theorem}
\newtheorem{lemma}{Lemma}

\newtheorem{definition}{Definition}

\newtheorem{remark}{Remark}
\title{\LARGE \bf Quantum Implemention of an LTI System with the Minimal 
Number of Additional Quantum Noise Inputs.
}

\author{Shanon L.~Vuglar, Ian R.~Petersen%
\thanks{Shanon L. Vuglar is with the School of Engineering and Information 
Technology, 
        University of New South Wales at the Australian Defence Force Academy, Canberra ACT 2600, Australia.
         {\tt\small shanonvuglar@vuglar.com} }%
\thanks{Ian R. Petersen is with the School of Information Technology and Electrical Engineering, 
        University of New South Wales at the Australian Defence Force Academy, Canberra ACT 2600, Australia.
         {\tt\small i.r.petersen@gmail.com} } }%

\begin{document}

\maketitle
\thispagestyle{empty}
\pagestyle{empty}

\begin{abstract}
Physical Realizability addresses the question of whether it is possible to 
implement a given LTI system as a quantum system. It is in general not true 
that a given synthesized quantum controller described by a set of stochastic 
differential equations is equivalent to some physically meaningful quantum 
system. However, if additional quantum noises are permitted in the 
implementation it is always possible to implement an arbitrary LTI system as a 
quantum system. In this paper we give an expression for the exact number of 
noises required to implement a given LTI system as a quantum system. 
Furthermore, we focus our attention on proving that this is 
a minimum, that is, it is not possible to implement the system as a quantum 
system with a smaller number of additional quantum noises. 
\end{abstract}

%%%%%%%%%%%%%%%%%%%%%%%%%%%%%%%%%%%%%%%%%%%%%%%%%%%%%%%%%%%%%%%%%%%%%%%%%%%%%%%%
\section{Introduction} \label{sec:intro}
As the collective research effort into applications that rely on quantum 
effects intensifies, the topic of quantum control becomes increasingly 
relevant \cite{JNP1,MaP4,GJ09,NJD09,NJP1,MAB08,
YAM06, WM10, VuP11a}. 
Such applications include quantum computing, quantum control and 
precision metrology. Quantum applications tend to be at the cutting edge 
of modern technology and have the potential to revolutionize the world we 
live in. However, their viability is dependent, at least in part, on quantum 
control. As an example, consider the recent results presented in \cite{VMS12}. Here, 
experimental success in maintaining a qubit in an oscillating superposition 
state through the use of feedback
control is highly relevant to the field of quantum computing.

Of particular interest is the sub-field of coherent quantum control 
\cite{JNP1,MaP4,NJP1,MAB08}. 
By coherent quantum control, we mean that both the plant and controller are 
quantum systems coupled together via some type of quantum mechanism. An example 
could be an optical system coupled to a second optical system specifically 
designed to control some aspect of the first system. This type of setup has 
several advantages. Firstly, and perhaps most importantly, it avoids direct 
measurement of the quantum state of the system, thus avoiding the collapse of 
the quantum state that inevitably occurs during observation (measurement) and 
the associated loss of information. This is particularly relevant to quantum 
computing where we wish to manipulate quantum states. Secondly, it may be that 
by implementing the controller with the same type of system as the plant they 
will be of similar time scales, and there may be advantages in terms of speed 
of the control mechanism as compared to implementing the controller via some 
other method and coupling it to the quantum system via measurement and 
actuation.

When considering coherent quantum control the question of physical 
realizability naturally arises; given a controller that has been designed 
by some standard means (eg. LQG or $H^\infty$ controller synthesis) is it 
possible to implement this controller as a quantum system? Unlike classical 
(i.e. not quantum) controllers which we may regard here as always being 
possible to implement at least approximately (but with arbitrary precision) 
via either analogue or digital electronics, it is in general not true that a 
given synthesized quantum controller described by a set of stochastic 
differential equations is equivalent to some physically meaningful quantum 
system. Several recent papers have addressed this issue of physical 
realizability \cite{JNP1,MaP3,VuP11a,VuP12a,VuP12b}, and of particular 
relevance is \cite{JNP1} in which 
the authors demonstrated that by incorporating additional quantum noises it is 
always possible to implement an arbitrary LTI system as a quantum system.

\begin{figure}[h]
	\includegraphics[trim = 0mm 0mm 0mm -10mm, scale=0.4]{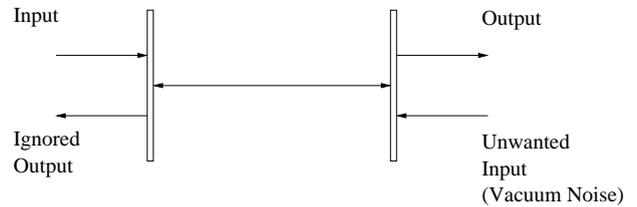}
	\caption{\label{fig} Optical Cavity}
\end{figure}

To better understand what is meant by additional quantum noises consider the 
following rather simplistic example: suppose that in implementing a quantum 
controller as an optical system, the controller design calls for a laser to be 
passed through an optical cavity as shown in Figure \ref{fig}. 
Here a naive approach would 
be to consider this device as having a single input and single output, however 
in fact there is a second input to the cavity: the mirror on the right which 
produces the output also allows causes the cavity to be coupled to a vacuum 
noise input; e.g. \cite{GZ00}. It turns out that to obtain correct results when 
modeling such a system it is imperative to take this additional vacuum 
noise source 
into account. It is in this way that additional quantum noise sources may 
inevitably enter the system when attempting to implement a controller as a 
quantum system; see also \cite{NJD09}.

In general, additional noise sources when implementing a 
controller are undesirable and previous work 
\cite{JNP1,VuP11a,VuP12a,VuP12b}, has gone some way to 
addressing the question of how many such noises are required in order to 
implement a given LTI system as a quantum system. In particular several 
previous results have provided upper and lower bounds on the number of 
additional quantum noises required as well as methods for determining when it is 
possible to implement a given quantum system with the additional noises equal 
to this lower bound. In this paper we provide a much stronger result; 
we give an expression for the exact number of noises required to implement an 
LTI system as a quantum system and focus our attention on proving that this is 
a minimum, that is, it is not possible to implement the system as a quantum 
system with a smaller number of additional quantum noises. 

The remainder of the paper proceeds as follows. In Section \ref{sec:model} 
we describe the quantum system model used throughout this paper. In Section
\ref {sec:realization} we define what is meant by \emph{physically realizable} 
and outline previous results. In Section \ref{sec:result} we present our 
main result. An example and our conclusion follow in Sections \ref{sec:ex} 
and \ref{sec:conc} respectively.

\section{Quantum System Model} \label{sec:model}
As in \cite{JNP1},
the linear quantum systems under consideration are assumed to be 
noncommutative stochastic systems described by quantum stochastic 
differential equations (QSDEs) of the form
\begin{eqnarray}
	\dif x(t) &=& A x(t) \dif t +
		\begin{bmatrix}B_1 & B\end{bmatrix}
		\begin{bmatrix}\dif v(t) \\ \dif u(t)\end{bmatrix};
			\qquad x(0) = x_0 \nonumber \\
	\dif y(t) &=& C x(t) \dif t +
	\begin{bmatrix}D_1 & 0_{n_y\times n_u}\end{bmatrix}
		\begin{bmatrix}\dif v(t) \\ \dif u(t)\end{bmatrix}
	\label{eqn:model}
\end{eqnarray}
where $x(t) = \begin{bmatrix}x_1(t) & \cdots & x_n(t)\end{bmatrix}^T$
is a column vector of $n$ self-adjoint system variables.
The noise $v(t) = \begin{bmatrix}v_1(t) & \cdots & v_{n_v}(t)
\end{bmatrix}^T$ is a vector of noncommutative Wiener processes 
(in vacuum states) with Ito products 
$\dif v(t) \dif v^T(t) = F_v \dif t$
where $F_{v}$ is non-negative Hermitian. 
$\dif u(t)$ is a column vector of signals of the form 
$\dif u(t) = \beta_u(t) \dif t + \dif \tilde{u}(t)$ where $\tilde{u}(t)$
is the noise part of $u(t)$ (with Ito products
$\dif \tilde{u}(t) \dif \tilde{u}^T(t) = F_{\tilde{u}} \dif t$
where $F_{\tilde{u}}$ is non-negative Hermitian)
and $\beta_u(t)$ is the adapted, self adjoint part of $u(t)$. $u(t)$  
represents the input to the system. $n,n_v,n_u$ and $n_y$ are all assumed to 
be even (this is because in the quantum harmonic oscillator the system 
variables always occur as conjugate pairs, see \cite{WM10}).
$A,B,C,B_1$ and $D_1$ 
are appropriately 
dimensioned real matrices describing the dynamics of the system. For further
details see \cite{JNP1}.

For simplicity we restrict our attention to the case where $n_y = n_u$.
\section{Physical Realizability} \label{sec:realization}
\subsection{Definition} \label{subsec:def}
As in \cite{JNP1,MaP3,MaP4,VuP11a}, 
the concept of \emph{physically realizable}
means that the system dynamics described by (\ref{eqn:model}) 
correspond to those of an open quantum harmonic oscillator. 

\begin{definition}
	By the \emph{canonical commutation relations} we mean that the system
	variables $x$ satisfy the commutation relations 
	$\begin{bmatrix}x_i(t),x_j(t)\end{bmatrix} = 2i\Theta_{ij}$
	where $\Theta$ is a block diagonal matrix with each diagonal block
	equal to $J = \left[ \begin{smallmatrix}0 & 1 \\ -1 & 0
	\end{smallmatrix} \right].$
\end{definition}

\begin{definition}
	By \emph{degenerate canonical commutation relations}, we mean that for 
	the system (\ref{eqn:model}), $\Theta$
	as defined above is a block diagonal matrix with
	a zero matrix block followed by diagonal block's 
	equal to $J$ as above, along the diagonal.
\end{definition}

The canonical commutation relations describe systems of a purely quantum nature
whereas the degenerate canonical commutation relations describe hybrid systems
consisting of both classical and quantum components.

\begin{definition}
	(See \cite [Definition 3.1]{JNP1}).
	The system described by (\ref{eqn:model}) 
	is an \emph{open quantum harmonic oscillator} if $\Theta$ is
	canonical and there exist a quadratic Hamiltonian $H = \frac{1}{2}
	x(0)^TRx(0)$,
	with a real, symmetric, $n \times n$ matrix $R$, and a coupling
	operator $L = \Lambda x(0$), with complex-valued $\frac{1}{2}
	\left( n_v + n_u
	\right)	\times n$ coupling matrix $\Lambda$, such that the matrices 
	$A, \begin{bmatrix}B_1 & B \end{bmatrix}, C$ and $\begin{bmatrix}
	D_1 & 0_{n_y \times n_u} \end{bmatrix}$ are given by:
	\begin{eqnarray}
		A &=& 2 \Theta \left(R + \mathfrak{Im}\left(
			\Lambda^{\dagger}\Lambda \right) \right);
			\label{eqn:a} \\
		\begin{bmatrix} B_1 & B \end{bmatrix}
			&=& 2i \Theta \begin{bmatrix}
			-\Lambda^{\dagger} & \Lambda^T \end{bmatrix}\Gamma; 
			\label{eqn:b} \\
		C &=& P^T \begin{bmatrix}
			\Sigma_{n_y} & 0 \\ 0 & \Sigma_{n_y} \end{bmatrix}
			\begin{bmatrix} \Lambda + \Lambda^\# \\
				-i\Lambda + i\Lambda^\# \end{bmatrix};
				\label{eqn:c} \\
		\begin{bmatrix} D_1 & 0_{n_y \times n_u} \end{bmatrix}
			&=&\begin{bmatrix} I_{n_y \times n_y} &
			0_{n_y \times \left(n_v + n_u - n_y\right)} 
			\end{bmatrix}.\label{eqn:d} 
	\end{eqnarray}
	Here: $\Gamma_{(n_v + n_u) \times (n_v + n_u)} = P \diag (M)$; 
		$M = \frac{1}{2}\left[ \begin{smallmatrix}1 & i \\ 1 & -i 
			\end{smallmatrix} \right]$;  
		$\Sigma_{n_y} = \begin{bmatrix}
			I_{\frac{1}{2}n_y \times \frac{1}{2}n_y} &
			0_{\frac{1}{2}n_y \times \frac{1}{2}\left(
				n_v + n_u - n_y \right) } \end{bmatrix}$;  
	$P$ is the appropriately dimensioned square permutation 
	matrix such that \linebreak
	$P \begin{bmatrix}a_1 & a_2 & \cdots & a_{2m} \end{bmatrix}
	=\begin{bmatrix}a_1 & a_3 & \cdots & a_{2m-1} 
	a_2 & a_4 & \cdots & a_{2m} \end{bmatrix}$ and 
	$\diag (M)$ is the appropriately dimensioned square block diagonal 
	matrix with the matrix $M$ occurring along the diagonal. (Note:  
	dimensions of $P$ and $\diag (M)$ can always be determined from the
	context in which they appear.) $\mathfrak{Im}\left(.\right)$ 
	denotes the imaginary part of a matrix, ${}^\#$ denotes the complex 
	conjugate of a matrix, and ${}^\dagger$ denotes the 
	complex conjugate transpose of a matrix.
\end{definition}

\begin{definition}
	(See \cite [Definition 3.3]{JNP1}).
	The system (\ref{eqn:model}) is said to be physically realizable if one 
of the following holds:
\begin{enumerate}
	\item $\Theta$ is canonical and (\ref{eqn:model}) represents 
		an open quantum harmonic oscillator.
	\item $\Theta$ is degenerate canonical and there exists an augmentation
		which, after a suitable relabeling of the components, 
		represents the dynamics of an open quantum harmonic oscillator
		\cite{JNP1}.
\end{enumerate}
\end{definition}

\begin{theorem}
	(see \cite [Theorem 3.4]{JNP1})
	The system (\ref{eqn:model}) is physically realizable if and only
	if:
	\begin{eqnarray}
		iA\Theta + i\Theta A^T + 
			\begin{bmatrix}B_1 & B \end{bmatrix} T_w
			\begin{bmatrix}B_1 & B \end{bmatrix} ^T &=& 0 
	\end{eqnarray}
	\begin{eqnarray}
		\begin{bmatrix}B_1 & B \end{bmatrix} 
		 \begin{bmatrix} I \\ 0 \end{bmatrix} &=& 
			\Theta C^T P^T
			\begin{bmatrix} 0 & I \\ -I & 0 \end{bmatrix} P 
				\nonumber \\
		&=& \Theta C^T \mbox{diag}(J)
		\label{eqn:realizable}
	\end{eqnarray}
	and $D_1$ satisfies (\ref{eqn:d}). Here 
	$$T_w = \frac{1}{2} \begin{bmatrix} 
		F_v - F_v{}^T & 0 \\ 0 & F_{\tilde{u}} - F_{\tilde{u}}{}^T 
	\end{bmatrix}.$$ 
\end{theorem}

In this paper we consider the problem of implementing a linear quantum system 
with a 
given transfer function as a fully quantum system and as such focus on the 
case where $\Theta$ is canonical.

\subsection{Previous Results} \label{subsec:previous}
In \cite{JNP1}, it was demonstrated that by incorporating additional 
quantum noises, an arbitrary linear time invariant system could always be 
physically realized as a quantum system. In particular, the following lemma 
relating to the physical realizability of a purely quantum controller with 
canonical commutation relations was proved.

\begin{lemma}
	(See \cite [Lemma 5.6]{JNP1}).
	Suppose $F_u = I + iJ$, $F_v = I + i J$,
	$A,B$ and $C$ are such that
	$A \in \mathbb{R}^{n\times n}, 
	B \in \mathbb{R}^{n\times n_u}, 
	C \in \mathbb{R}^{n_y\times n},$
	and $\Theta = \diag(J)$ is canonical. Then there exists an even 
	integer	$n_v \ge n_y$ and matrix $B_1 \in \mathbb{R}^{n\times n_v}$
	such that the system (\ref{eqn:model}) is physically realizable.
\end{lemma}

%New
\begin{remark}
By \emph{minimal additional noises} we mean that $n_v = n_y$. It follows 
from \cite [Theorem 3.4]{JNP1} that the number of outputs $n_y$ is a 
lower bound on the number of additional noises $n_v$, necessary for 
physical realizability for a system described by a strictly proper 
transfer function.
\end{remark}
%end new

In general, the incorporation of additional noises is undesirable, and in 
\cite{VuP11a} the question of how many additional noises are required to 
implement a strictly proper LTI Quantum System was addressed. Drawing on the 
results 
contained in \cite{JNP1}, in \cite{VuP11a} an upper bound on the number of 
additional noises required was given. Further to this, a condition in terms 
of a certain non-standard Riccati equation was obtained 
for when a given transfer function (rather than a specific state space 
realization) could be physically realized using only a minimum number of 
additional 
quantum noises. The main results of \cite{VuP11a} are reproduced here for 
the sake of completeness.

\begin{theorem}
	\label{thm:1}
	(See \cite [Theorem 1]{VuP11a}). Consider an LTI system of the form (\ref{eqn:model}) where 
	$A,B$ and $C$ are given and the system commutation matrix $\Theta$ is
	canonical. There exists $B_1$ and $D_1$ such that the system is 
	physically 
	realizable with the number of quantum noises in $\dif v$ 
	equal to $n_u + 2 \left(n - n_{\lambda}\right)$ where $n_\lambda$ 
	is the multiplicity of the least (i.e. most negative) eigenvalue of
	the matrix 
	$i \left( \Theta B \Theta B^T \Theta - \Theta A - A^T \Theta
	- C^T \Theta C \right)$.
\end{theorem}

In the special case that $n_{\lambda} = n$,
it follows that $n_v = n_u$, which is the 
minimum number of quantum noises which need to be added for 
physically realizability.

\section{Main Result} \label{sec:result}

\begin{theorem}
	\label{thm:3}
	Consider an LTI system of the form (\ref{eqn:model}) where 
	$A,B$ and $C$ are given. 
	There exists $B_1$ and $D_1$ such that the system is physically 
	realizable with canonical commutation matrix $\Theta$, and with 
	the number of additional quantum noises $n_v$ 
	equal to $n_u + r$ where $r$ 
	is the rank of the matrix 
	$\left( \Theta B \Theta B^T \Theta - \Theta A - A^T \Theta
	- C^T \Theta C \right)$. 
	Conversely, suppose there exists $B_1$ and $D_1$ such that the system 
	(\ref{eqn:model}) where	$A,B$ and $C$ are given is physically 
	realizable with canonical commutation matrix $\Theta$ and the number 
	of additional quantum noises equal to $n_v$. 
	Then $n_v \ge n_u + r$ where $r$ is the rank of the matrix 
	$\left( \Theta B \Theta B^T \Theta - \Theta A - A^T \Theta
	- C^T \Theta C \right)$. 
	That is, it is not possible to choose 
	$B_1$ and $D_1$ such that the system is physically realizable and 
	the number of additional quantum noises $n_v$ is 
	less than $n_u + r$.
\end{theorem}

\begin{remark}
Theorem 3 is stronger than Theorem 2 for two reasons: firstly, in general it 
will give a better result in terms of the number of additional quantum noises 
sufficient for physical realizability, and secondly, Theorem 2 does not address 
the necessity of these noises whereas Theorem 3 does.
\end{remark}

\begin{remark}
Theorem 2 and Theorem 3 are consistent with respect to the special case where 
the minimum number of additional quantum noises $n_v = n_u$ are sufficient 
(and necessary) for physical realizability. It will be shown in the proof that 
follows that 
	$\left( \Theta B \Theta B^T \Theta - \Theta A - A^T \Theta
	- C^T \Theta C \right)$
has eigenvalues that occur in $+/-$ pairs. As such, the case in Thereom 2 
where $n_{\lambda} = n$ , (that is all the eigenvalues are the same,) can only
only occur when all the eigenvalues are identically zero which corresponds to 
the case in Theorem 3 where $r = 0$, (that is,  
	$ \Theta B \Theta B^T \Theta - \Theta A - A^T \Theta
	- C^T \Theta C = 0$).
\end{remark}

\begin{proof}
The proof is structured as followed: we first show that $n_v = n_u + r$ 
additional 
noises are sufficient for physical realizability; we then show that 
$n_v \ge n_u + r$ additional noises are necessary.
	
In \cite{JNP1} a method was given to construct matrices $R, \Lambda , B_1$ and 
$D_1$ in (\ref{eqn:a}) - (\ref{eqn:d}) such that $A,B$ and $C$ were realized 
as required. Specifically the construction given in \cite{JNP1} is as follows: 
	\begin{eqnarray*}
		D_1 &=&\begin{bmatrix} I_{n_y \times n_y} &
			0_{n_y \times \left(n_v - n_y\right)} 
			\end{bmatrix}; \\
		R &=&  -\frac{1}{4}\left( \Theta A + A^T \Theta^T \right); \\
		B_1 &=& \begin{bmatrix}B_{1,1} & B_{1,2} \end{bmatrix};\\
		\Lambda &=& \begin{bmatrix}
			\frac{1}{2}C^T P^T \begin{bmatrix}
				I \\ iI \end{bmatrix} &
			\Lambda^T_{b1} & \Lambda^T_{b2} \end{bmatrix}^T; \\
		B_{1,1} &=& \Theta C^T \diag (J); \\
		\Lambda_{b2} &=& -i \begin{bmatrix} I &
			0 \end{bmatrix}
			P \diag (M) \nonumber B^T \Theta; \\
		B_{1,2} &=& 2i\Theta \begin{bmatrix}
			-\Lambda^{\dagger}_{b1} & \Lambda^T_{b1} \end{bmatrix}
			P \nonumber \diag (M)
	\end{eqnarray*}
	where $n_v \ge n_u + 2$.
	Here, $\Lambda_{b1}$ is any complex $\frac{1}{2}\left(n_v - n_u\right)
	\times n$ matrix such that
	\begin{eqnarray}
		\Lambda_{b1}^{\dagger}\Lambda_{b1} &=&  \Xi_1  
		+ i \left( \begin{array}{c} 
			\frac{A^T \Theta^T  - \Theta A}{4} \\% {} \\
			\scriptstyle{ {} - \frac{1}{4}C^T P^T
			\left[ \begin{smallmatrix} 0 & I \\ 
				-I & 0  \end{smallmatrix} \right] PC} \\% {} \\
		\scriptstyle{ {} - \mathfrak{Im}
		\left( \Lambda^{\dagger}_{b2} \Lambda_{b2} \right)} 
		\end{array} \right) 
		%\bigg( 
		%\frac{A^T \Theta^T  - \Theta A}{4} - \frac{1}{4}C^T P^T
		%\begin{bmatrix}0 & I \\ -I & 0 \end{bmatrix} PC \nonumber \\ 
		%&&{} - \mathfrak{Im}
		%\left( \Lambda^{\dagger}_{b2} \Lambda_{b2} \right) 
		%\bigg) 
		\label{eqn:XX}
	\end{eqnarray}
	where $\Xi_1$ is any real symmetric $n \times n$ matrix such that
	$\Lambda_{b1}^{\dagger}\Lambda_{b1}$ is nonnegative definite.

$\Lambda_{b1}$ is constructed as follows: a real 
symmetric $n \times n$ matrix $\Xi_1$ is first constructed such that
the r.h.s. of (\ref{eqn:XX}) is nonnegative definite.
$\Lambda_{b1}$ is then constructed such that (\ref{eqn:XX}) holds. 

Note: the construction of $\Lambda_{b1}$ is of particular interest 
as $\Lambda_{b1}$ has precisely $\frac{1}{2}\left(n_v - n_u
\right)$ rows and thus
determines the number of additional quantum noises required in this 
implementation.

	Note: The construction above (as given in \cite{JNP1}) only allows for 
	$n_v \ge n_u + 2$. However in the special case that the imaginary part 
	of the r.h.s. of (\ref{eqn:XX}) is precisely zero, $\Xi_1$ can be chosen 
	to be zero, $\Lambda_{b1}$ and $B_{1,2}$ vanish, and $n_v = n_u$.

We now provide a method for choosing $\Xi_1$ and 
$\Lambda_{b1}$ to obtain the result.

Note: 
It is desirable to construct $\Xi_1$ such that the
$\Lambda_{b1}^{\dagger}\Lambda_{b1}$ 
is of minimum rank. This will allow $\Lambda_{b1}$ to be constructed 
with minimum rows, thus the additional quantum noises required in this 
implementation, will be the minimum possible under this method of constructing 
$R, \Lambda , B_1$ and $D_1$.

In \cite{VuP11a} we show that (\ref{eqn:XX}) can be rewritten as
$$\Xi_2 =  \Xi_1 + \frac{i}{4} \tilde{S}$$
where 
\begin{eqnarray*}
\Xi_2 &=&  \Lambda_{b1}^{\dagger}\Lambda_{b1}; \\ 
\tilde{S} &=& \Theta B \Theta B^T \Theta - A^T \Theta - \Theta A
	- C^T \Theta C .
\end{eqnarray*}

Note that the matrix $\tilde{S}$
is real, skew symmetric. Thus
$ S = \frac{i}{4} \tilde{S} $
is hermitian, has real eigenvalues and is diagonalizable: 
$S = U^{\dagger}DU$ where $D$ is diagonal and $U$ is unitary. 

We wish to find a real, symmetric $\Xi_1$ such that 
$\Xi_1 + S$ is positive semi-definite and of minimum rank.
Let $\Xi_1 = U^\dagger \left| D \right| U$ where 
$\left| D \right|$ is the diagonal matrix with values equal to the absolute 
values of the corresponding entries in $D$. Then $\Xi_1$ is real, 
symmetric and $\Xi_2 = \Xi_1 + S \ge 0$. Further, $\Xi_2$ has rank equal to 
half that of $S$ and this is the minimum rank possible for all allowed 
choices of $\Xi_1$.

First we show that $\Xi_1$ so obtained is real and symmetric. 
Observe that $\Xi_1 = {\Xi_1}^\dagger; \Xi_1 \ge 0$.
Also: 
$${\Xi_1}^2 = U^\dagger {\left| D \right|}^2 U = U^\dagger D^2 U = S^2$$
Here, $S$ is purely imaginary, thus $S^2$ is real and ${\Xi_1}^2 \ge 0$ is also 
real, and therefore has a real square root. From the 
uniqueness of the positive semi-definite square root of a positive 
semi-definite matrix 
\cite [Theorem 7.2.6] {HJ85}, we conclude that $\Xi_1$ is real.
Further, since $\Xi_1$ is Hermitian, $\Xi_1$ is symmetric.

We now show the rank condition. Observe the following about the eigenvalues of 
$S$: $S$ is hermitian so its eigenvalues are real, thus the 
eigenvalues of $\tilde{S}$ are purely imaginary, but since $\tilde{S}$ 
is real its eigenvalues 
occur in complex conjugate pairs. Thus $D$ is of the form:
$$D = \begin{bmatrix} 
	\lambda_1 & 0 & 0 & 0 & \cdots \\
	0 & -\lambda_1 & 0 & 0 & \cdots \\
	0 & 0 & \lambda_2 & 0 & \cdots \\
	0 & 0 & 0 & -\lambda_2 & \cdots \\
	\vdots & & & & \ddots
\end{bmatrix}; \lambda_i \ge 0, \forall i;$$

$$\left| D \right| = \begin{bmatrix}
	\lambda_1 & 0 & 0 & 0 & \cdots \\
	0 & \lambda_1 & 0 & 0 & \cdots \\
	0 & 0 & \lambda_2 & 0 & \cdots \\
	0 & 0 & 0 & \lambda_2 & \cdots \\
	\vdots & & & & \ddots
\end{bmatrix}; \lambda_i \ge 0, \forall i;$$
$$\left| D \right| + D = \begin{bmatrix}
	0 & 0 & 0 & 0 & \cdots \\
	0 & \lambda_1 & 0 & 0 & \cdots \\
	0 & 0 & 0 & 0 & \cdots \\
	0 & 0 & 0 & \lambda_2 & \cdots \\
	\vdots & & & & \ddots
\end{bmatrix}; \lambda_i \ge 0, \forall i.$$
From this, it can be seen that $\left| D \right| + D$ has rank, half that of 
$D$. Since
\begin{eqnarray*}
	\Xi_2 &=& \Xi_1 + S \\
		&=& U^\dagger \left| D \right| U + U^\dagger D U \\
		&=& U^\dagger (\left| D \right| + D) U,
\end{eqnarray*}
it follows that $\Xi_2$ has rank, half that of $S$. 

Since $S$ and $\tilde{S}$ 
have the same rank, $\Xi_2$ has rank $\frac{r}{2}$ where $r$ the 
rank of $\tilde{S}$. Since $\Xi_2 \ge 0$ has rank $\frac{r}{2}$, it is 
possible to construct $\Lambda_{b1}$ with $\frac{r}{2}$ rows, 
such that $\Xi_2 = \Lambda_{b1}^\dagger \Lambda_{b1}$. Recall that, 
$\Lambda_{b1}$ has precisely $\frac{1}{2}\left(n_v - n_u \right)$ rows, and we 
have $n_v = n_u + r$, that is, 
the system is
physically 
realizable with the number of additional quantum noises $n_v$ 
equal to $n_u + r$ where $r$ 
is the rank of the matrix 
$\tilde{S} = \left( \Theta B \Theta B^T \Theta - \Theta A - A^T \Theta
- C^T \Theta C \right)$.

We now consider the second part of the theorem and show that $n_v \ge n_u + r$
additional noises are necessary. To do so it is sufficient to show that the 
number of columns of 
$B_1 = \begin{bmatrix}B_{1,1} & B_{1,2} \end{bmatrix}$ is greater than or 
	equal to $n_u + r$.

We consider the dimensions of $B_{1,1}$ first. From (\ref{eqn:realizable}) 
we obtain
$$\begin{bmatrix} B_{1,1} & B_{1,2} & B \end{bmatrix} 
	\begin{bmatrix} I \\ 0 \end{bmatrix} =
	\Theta C^T diag(J).$$
By observing the dimensions of matrices on the l.h.s. this reduces to:
$$ B_{1,1} = \Theta C^T diag(J),$$
Observe here that $B_{1,1}$ has precisely as many columns as $C$ has 
rows, that is, $B_{1,1}$ has $n_y = n_u$ columns.

We next consider the dimensions of $B_{1,2}$. 
From (\ref{eqn:b}) it can be shown that:
\begin{eqnarray}
	B &=&  2i \Theta \begin{bmatrix} -\Lambda^\dagger_{b2} &
		\Lambda^T_{b2} \end{bmatrix} \Gamma; \\
	B_{{1,1}} &=&  2i \Theta \begin{bmatrix} -\Lambda^\dagger_{b0} &
		\Lambda^T_{b0} \end{bmatrix} \Gamma; \label{eqn:b11}\\
	B_{{1,2}} &=&  2i \Theta \begin{bmatrix} -\Lambda^\dagger_{b1} &
		\Lambda^T_{b1} \end{bmatrix} \Gamma;
\end{eqnarray}
where $$\Lambda = \begin{bmatrix} \Lambda_{b0} \\ 
	\Lambda_{b1} \\ \Lambda_{b2} \end{bmatrix}.$$ 
That is, $B_{1,2}$ has twice the number of columns as $\Lambda_{b1}$ has 
rows. We wish to show therefore, 
that $\Lambda_{b1}$ has at least $\frac{r}{2}$ 
rows.

Consider,
$$
\mathfrak{Im}(\Lambda^\dagger \Lambda) =
\mathfrak{Im}(\Lambda^\dagger_{b0} \Lambda_{b0} ) +
\mathfrak{Im}(\Lambda^\dagger_{b1} \Lambda_{b1} ) +
\mathfrak{Im}(\Lambda^\dagger_{b2} \Lambda_{b2} ).$$
That is, 
$$
\mathfrak{Im}(\Lambda^\dagger_{b1} \Lambda_{b1} ) =
\mathfrak{Im}(\Lambda^\dagger \Lambda) -
\mathfrak{Im}(\Lambda^\dagger_{b0} \Lambda_{b0} ) -
\mathfrak{Im}(\Lambda^\dagger_{b2} \Lambda_{b2} ).$$

Rearranging (\ref{eqn:a}):
$$\frac{1}{2}\Theta^{-1}A = R + \mathfrak{Im}(\Lambda^\dagger \Lambda),$$
where $R$ and $\mathfrak{Im}(\Lambda^\dagger \Lambda),$ are the symmetric and 
skew-symmetric parts respectively of the left hand side. 
From this it can be shown that 
$$\mathfrak{Im}(\Lambda^\dagger \Lambda) = -\frac{1}{4}(\Theta A + A^T 
\Theta).$$

%%%%%%%%%%%%%%%%%%%%%%%%%%%%%%%%%%%%%%%%%%%%%%%%
Also, it is straightforward to verify that
$$\mathfrak{Im}(\Lambda^\dagger_{b0} \Lambda_{b0}) = 
\frac{1}{4}(C^T \Theta C),$$ and

$$\mathfrak{Im}(\Lambda^\dagger_{b2} \Lambda_{b2}) = 
-\frac{1}{4}(\Theta B \Theta B^T \Theta).$$

Substituting: 
\begin{eqnarray*}
	\lefteqn{i \times \mathfrak{Im}(\Lambda^\dagger_{b1} \Lambda_{b1} )} \\ 
	&=& 
	\frac{i}{4} \left(\Theta B \Theta B^T \Theta 
	- A^T \Theta - \Theta A	- C^T \Theta C \right) \\
	&=& \frac{i}{4}\tilde{S} = S
	.
\end{eqnarray*}

That is, 
$$\Lambda^\dagger_{b1} \Lambda_{b1} = \Xi_1 + \frac{i}{4}\tilde{S},$$
where $\Xi$ is the real part of $\Lambda^\dagger_{b1} \Lambda_{b1}$. 

Consider the following fact \cite [Fact 2.17.3] {BER05}: Let 
$A, B \in \mathbb{R}^{n \times m}$, then
$$\mbox{rank} (A + jB) = \frac{1}{2} \mbox{rank} 
\begin{bmatrix} A & B \\ -B & A \end{bmatrix}.$$

From this observe that 
\begin{eqnarray*}
	\mbox{rank} \left( \Lambda^\dagger_{b1} \Lambda_{b1} \right)
	&=& \mbox{rank} \left( \Xi_1 + i \frac {\tilde{S}}{4} \right) \\
	&=& \frac{1}{2} \mbox{rank} 
\begin{bmatrix} \Xi_1 & \frac {\tilde{S}}{4}  \\ 
	- \frac {\tilde{S}}{4} & \Xi_1 \end{bmatrix} \\
&\ge& \frac{1}{2} \mbox{rank} 
\begin{bmatrix} \Xi_1 & \frac {\tilde{S}}{4} \end{bmatrix} \\
&\ge& \frac{1}{2} \mbox{rank} \frac {\tilde{S}}{4}
\end{eqnarray*}

That is, independent of $\Xi_1$, 
$$\mbox{rank} \left(
\Lambda^\dagger_{b1} \Lambda_{b1} \right) \ge \frac{1}{2}\mbox{rank}S.$$ 

This in turn implies that $\Lambda_{b1}$ has at least $\frac{r}{2}$ rows, 
where, as before, $r$ is the rank of the matrix 
$\tilde{S} = \left( \Theta B \Theta B^T \Theta - \Theta A - A^T \Theta
- C^T \Theta C \right)$. But from (\ref{eqn:b11}), $B_{1,2}$ has twice as many 
columns as $\Lambda_{b1}$ has rows. 
That is, $B_{1,2}$ has at least $r$ columns and hence $B_1$ has at least 
$n_u + r$ columns. As such, the number of additional noises $n_v$ is greater 
than or equal to $n_u + r$. This concludes the proof of the theorem.
\end{proof}

\section{Illustrative Example}\label{sec:ex}
In this section we consider a system from \cite [Section VII,D]{JNP1}, in 
which an example of classical-quantum controller synthesis is given,
where the controller is implemented as a degenerate canonical controller with 
both classical and quantum degrees of freedom. Previously, in 
\cite{VuP12a} it was shown that this system can be physically 
realized as a fully quantum system with $n_v = 8$ additional quantum
noises. Here we apply our main result to show that this system can be 
implemented as a quantum system with only $n_v = 6$ additional quantum 
noises. Furthermore, it is not possible to implement this system as quantum 
system with less than $n_v = 6$ additional quantum noises. 

Consider a system of the form (\ref{eqn:model}) with 
\begin{eqnarray*}
A &=&  \left[ \begin{smallmatrix} -1.3894 \, I_{2 \times 2} &
		-0.4472 \, I_{2 \times 2} \\
		-0.2 \, I_{2 \times 2} &
		-0.25 \, I_{2 \times 2} \end{smallmatrix} \right]; \\
B &=&  \left[ \begin{smallmatrix} -0.4472 \, I_{2 \times 2} \\
		0_{2 \times 2} \end{smallmatrix} \right]; \\
C &=&  \left[ \begin{smallmatrix} -0.4472 \, I_{2 \times 2} &
		0_{2 \times 2} \end{smallmatrix} \right]; \\
n &=&  4; \quad n_u = 2; \quad \mbox{and} \quad n_y = 2.
\end{eqnarray*}
Applying \emph{Theorem 3}:
\begin{eqnarray*}
\tilde{S} &=& \left( \Theta B \Theta B^T \Theta - \Theta A - A^T \Theta
	- C^T \Theta C \right) \\
&=&  \left[ \begin{smallmatrix} 
0		& 2.3788	& 0		& 0.6472 \\
-2.3788		& 0		& -0.6472	& 0 \\
0		& 0.6472	& 0		& 0.5 \\
-0.6472		& 0		& -0.5		& 0
\end{smallmatrix} \right]
\end{eqnarray*}
has rank $r = 4$, therefore the system is physically realizable with 
$n_v = n_u + r = 2 + 4 = 6$ additional quantum noises. Furthermore, it is not 
possible to physically realize this system with less than $n_v = 6$ 
additional quantum noises. 

Remark: In \cite{VuP12a} we show that if we are only concerned with 
implementing the transfer function described by $\left\{A, B, C \right\}$ in 
this example then it is possible to implement a different system 
$\left\{\tilde{A}, \tilde{B}, \tilde{C} \right\}$ 
with the same transfer function and only 2 additional noises.

\section{Conclusion} \label{sec:conc}
Physical Realizability is particularly pertinent to coherent quantum control 
where we wish to implement a given synthesized controller as a quantum system. 
By incorporating additional quantum noises in the implementation it is always 
possible to make a given LTI system physically realizable, however 
incorporating additional quantum noises is undesirable. In this paper we have 
given an expression for the number of additional quantum noises that are 
necessary to make a given LTI system physically realizable. Furthermore, 
we have shown that is not possible to make the given system physically 
realizable with a smaller number of additional quantum noises.

\bibliography{irpnew}  
\bibliographystyle{IEEEtran}

\end{document}